\documentclass[a4paper]{jpconf}
\usepackage{graphicx}
\usepackage[pdf]{pstricks}
\usepackage{epstopdf}
\usepackage[square,sort&compress]{natbib}

\begin{document}

\title{A tool to convert CAD models for importation into Geant4}

\author{Carl Vuosalo, Duncan Carlsmith, Sridhara Dasu \newline
and Kimberly Palladino, \newline
on behalf of the LUX-ZEPLIN collaboration}

\address{University of Wisconsin-Madison}

\ead{covuosalo@wisc.edu}

\begin{abstract}
The engineering design of a particle detector is usually performed in a
Computer Aided Design (CAD) program, and simulation of the detector's performance
can be done with a Geant4-based program. However, transferring the detector
design from the CAD program to Geant4 can be laborious and error-prone.

SW2GDML is a tool that reads a design in the popular SOLIDWORKS CAD
program and outputs Geometry Description Markup Language (GDML), used
by Geant4 for importing and exporting detector geometries.
Other methods for outputting CAD designs are available, such as the STEP
format, and tools exist to convert these formats into GDML.
However, these conversion methods produce very large and unwieldy designs
composed of tessellated solids that can reduce Geant4 performance. In
contrast, SW2GDML produces compact, human-readable GDML that employs standard
geometric shapes rather than tessellated solids.

This paper will describe the development and current capabilities of SW2GDML
and plans for its enhancement. The aim of this tool is to automate
importation of detector engineering models into Geant4-based simulation
programs to support rapid, iterative cycles of detector design, simulation, and
optimization.
\end{abstract}

\section{Introduction}
The development of a device such as a sophisticated particle detector requires both engineering design and
simulation of performance under exposure to
particle flux. Design is performed in a Computer Aided Design (CAD) program like SOLIDWORKS~\cite{solidworks},
which is a long-standing market leader in the CAD industry. Simulation is typically
performed with Geant4 \cite{Agostinelli2003250, Allison2016186, 1610988},
which is extensively used in particle physics. The model of the device must be represented in both the CAD program
and in the simulation. Any discrepancies between these two models could lead to incorrect results and, ultimately, a faulty design of the device.

Transfer of the design from the CAD program to Geant4 is often done manually, which is a slow, laborious, and
error-prone process. As an alternative, many CAD
programs can output a design in a standard format, like STEP (STandard for the Exchange of Product model data)~\cite{ISO10303}.
From STEP, it is necessary to convert into the Geometry Description Markup Language (GDML) format used by Geant4 for importing
and exporting models. Tools like FASTRAD~\cite{trad} can convert STEP into GDML.
However, this method drops material properties and produces very large and unwieldy designs
composed of tessellated solids that can reduce Geant4 performance.

SW2GDML is a tool that reads a design in SOLIDWORKS and outputs GDML in a compact, human-readable format that uses standard 
geometric solids rather than tessellated shapes. It also includes the material composition of parts as specified in SOLIDWORKS in the converted GDML.
The converted design can be used in Geant4 without the performance penalties of tessellated forms that have large numbers of vertices.

\begin{figure}[htpb]
\begin{center}
\includegraphics[trim={0 1.0cm 0 11.0cm},clip,scale=0.4]{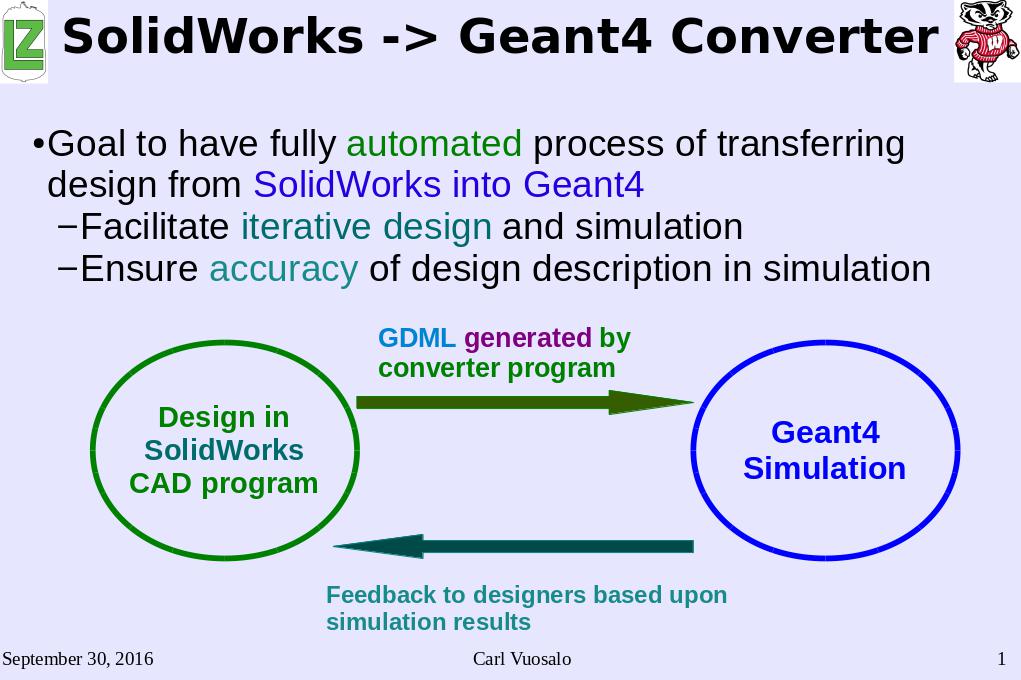}
\end{center}
\caption{\label{dcycle}Iterative design cycle for development of a model using a CAD program and Geant4 for simulation and evaluation of the physics
capabilities of the design.}
\end{figure}

With SW2GDML, a rapid, iterative design cycle becomes feasible. A model can be created in SOLIDWORKS, converted and imported into Geant4, and then
simulations can be run. The results of the simulations may show deficiencies in the design, which can be rectified in the SOLIDWORKS model. Then the
cycle can be repeated, as often as necessary, with the rapid, automated conversion from SOLIDWORKS to Geant4 the key to
speeding up the process (Fig.~\ref{dcycle}).

\section{Initial Development of SW2GDML}
\begin{figure}[htpb]
\begin{center}
\includegraphics[trim={1.0cm 1.5cm 0 3.0cm},clip,scale=0.6]{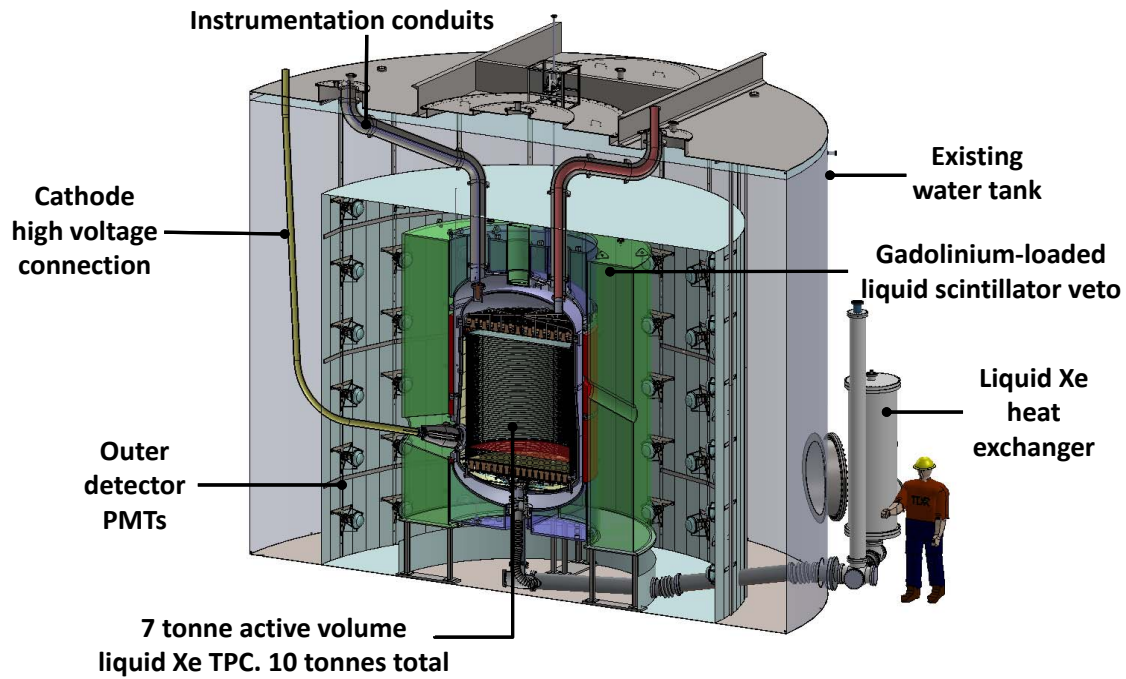}
\end{center}
\caption{\label{lzfullcolor}LZ detector overview. The inner detector will hold a 7-tonne fiducial volume of liquid xenon
for direct detection of dark matter. The outer detector serves to help characterize the low-background environment and veto background events.}
\end{figure}

\begin{figure}[htpb]
\begin{center}
\includegraphics[trim={4.0cm 1.0cm 4.0cm 1.0cm},clip,scale=0.6]{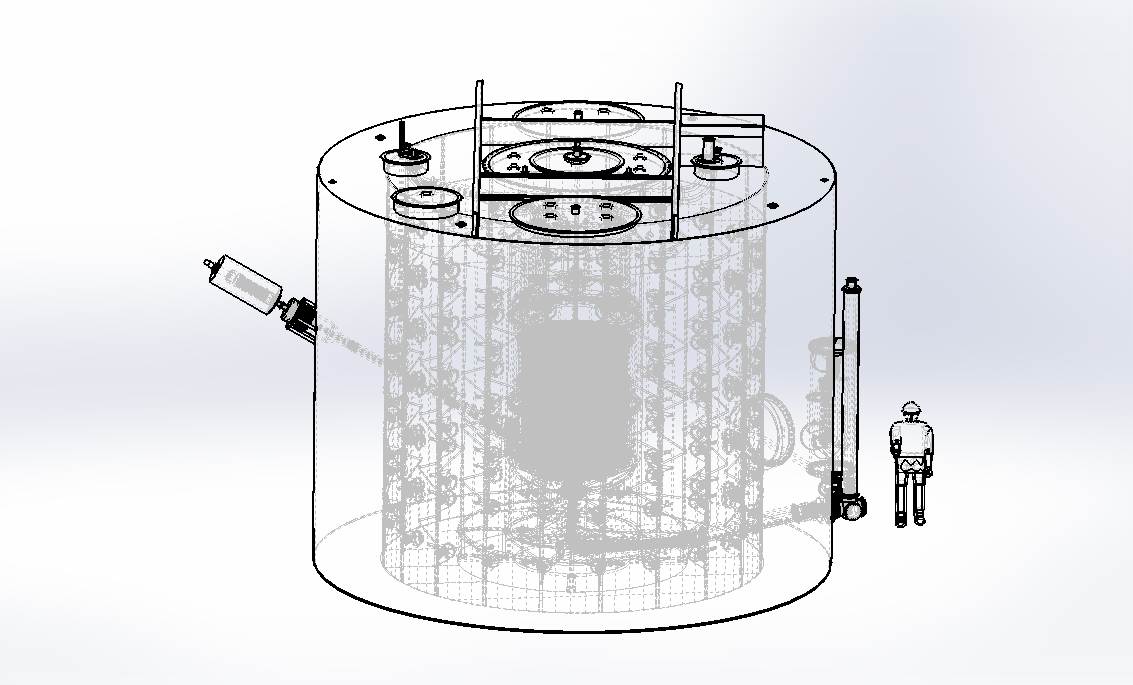}
\end{center}
\caption{\label{lzfullxray} Previous version of the LZ detector design, highlighting the outer detector photomultiplier tubes.}
\end{figure}

\begin{figure}[htpb]
\begin{center}
\includegraphics[trim={0 1.0cm 0 2.0cm},clip,scale=0.6]{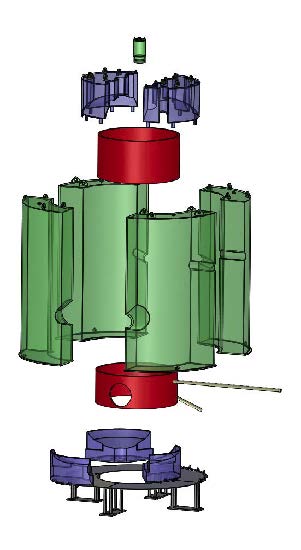}
\end{center}
\caption{\label{outerdet} LZ outer detector liquid scintillator tanks.}
\end{figure}

SW2GDML was initially developed for use in the design of the LUX-ZEPLIN (LZ) Dark Matter Experiment \cite{Akerib:2015cja}. The LZ collaboration is
building a detector that will hold 7 tonnes of liquid xenon and which will be installed
at the Sanford Underground Research Facility in the former Homestake gold mine in Lead, South Dakota.
When the LZ detector is installed and begins collecting data, it will be the most sensitive liquid xenon dark matter detector
to date.
Major components of the detector are being designed with SOLIDWORKS. Figure~\ref{lzfullcolor} shows an overview of the detector, Fig.~\ref{lzfullxray}
highlights the outer detector photomultiplier tubes, and Fig.~\ref{outerdet} shows the outer detector liquid scintillator tanks.
The task of converting key parts of the detector, like the liquid scintillator tanks, guided the early development of SW2GDML.

Divergence of the development time lines for SW2GDML and the LZ detector design, along with sparse resources for SW2GDML development,
limited the contribution of SW2GDML to the LZ design effort, but now the tool is available for use with other projects that employ
SOLIDWORKS and Geant4.

\section{Capabilities of SW2GDML}

\begin{figure}[htpb]
\begin{minipage}{16pc}
\includegraphics[scale=0.53,trim={9.0cm 2.5cm 9.0cm 2.0cm},clip,angle=100,origin=c]{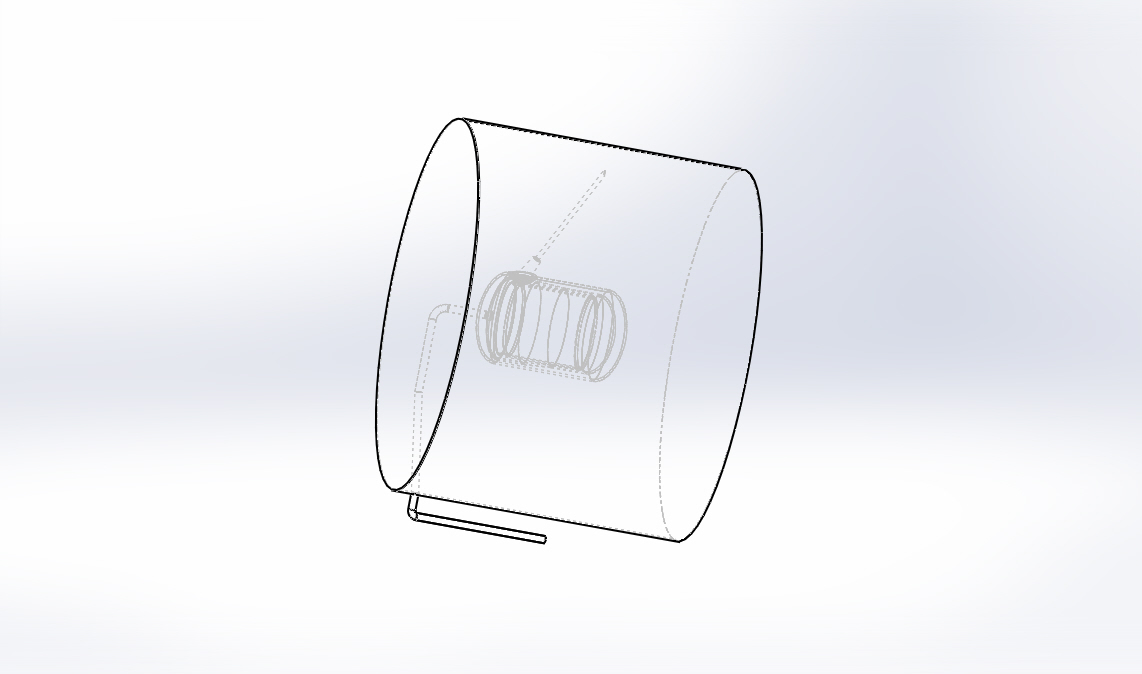}
\end{minipage}\hspace{2pc}%
\begin{minipage}{16pc}
\includegraphics[width=16pc,scale=0.1,angle=90,origin=c]{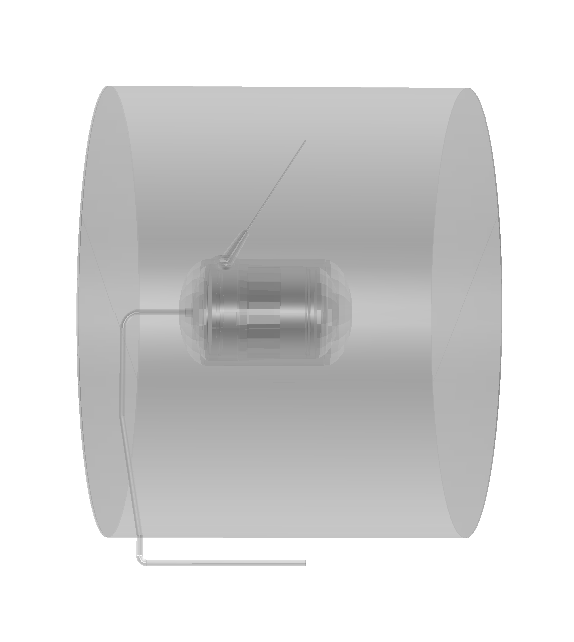}
\end{minipage} 
\caption{\label{swbigtank} Simple model of LZ outer and inner tanks, shown in SOLIDWORKS on the left, and automatically converted into Geant4 on the right.}
\end{figure}

\begin{figure}[htpb]
\begin{minipage}{16pc}
\includegraphics[scale=0.292]{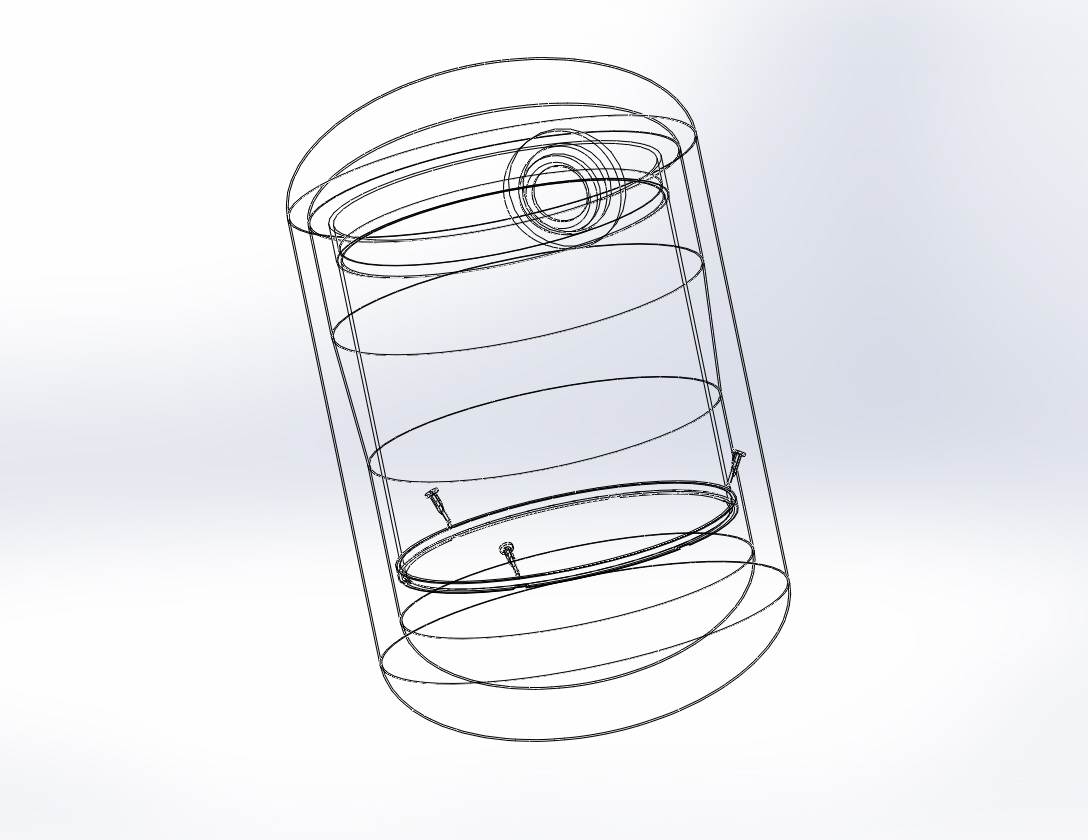}
\end{minipage}\hspace{2pc}%
\begin{minipage}{16pc}
\includegraphics[width=16pc,trim={6.0cm 5.0cm 10.0cm 3.0cm},clip,scale=0.2,angle=0,origin=c]{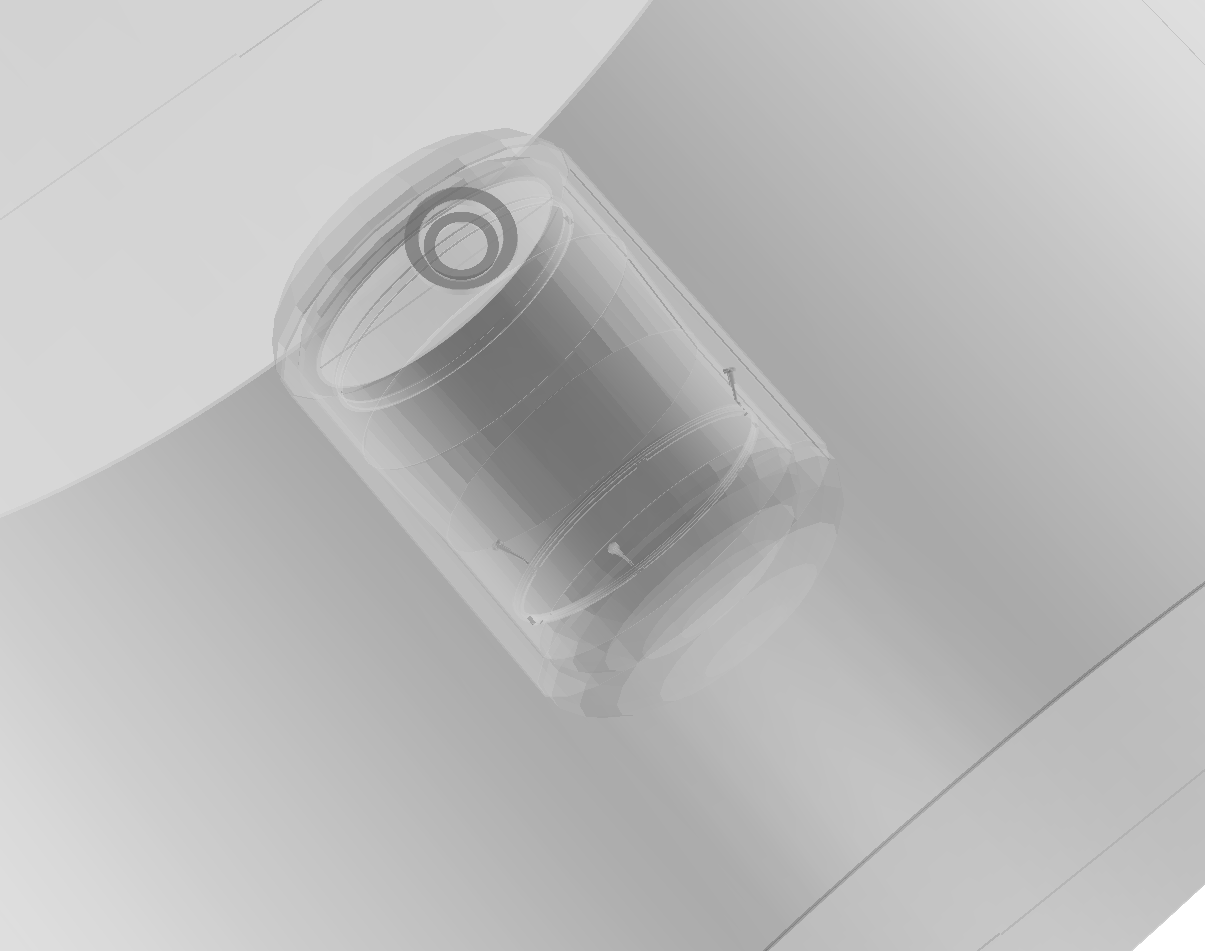}
\end{minipage} 
\caption{\label{swinnertank} Simple model of the LZ inner tank, shown in SOLIDWORKS on the left, and automatically converted into Geant4 on the right.}
\end{figure}

\begin{figure}[htpb]
\begin{minipage}{16pc}
\begin{center}
\includegraphics[trim={10.0cm 0 8.0cm 0},clip,scale=0.6]{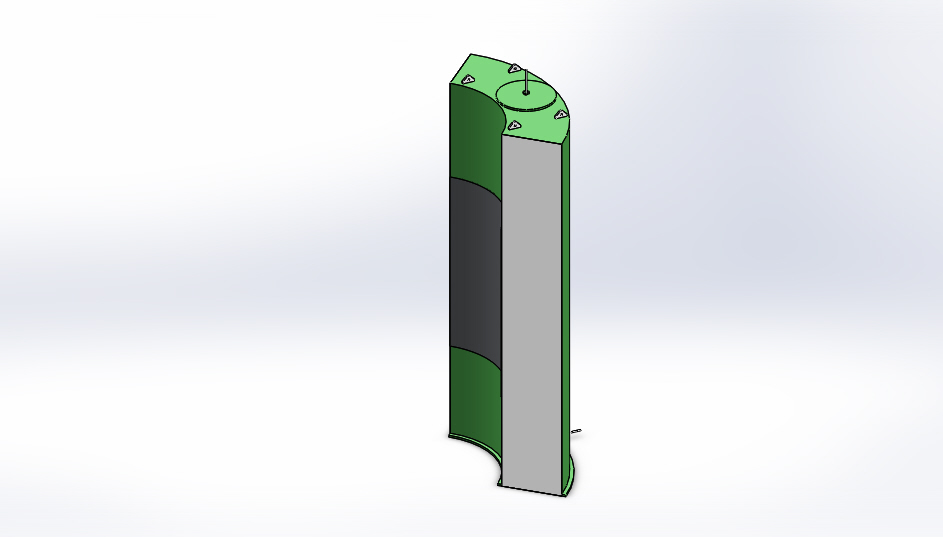}
\end{center}
\end{minipage}\hspace{2pc}%
\begin{minipage}{16pc}
\includegraphics[width=16pc,scale=0.1]{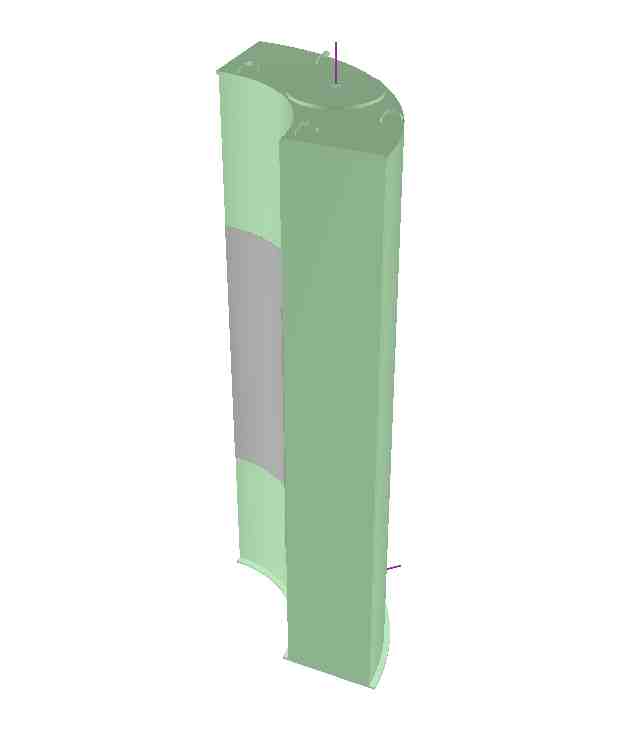}
\end{minipage} 
\caption{\label{swfulltank} Simple model of one LZ liquid scintillator tank, shown in SOLIDWORKS on the left, and automatically converted into Geant4 on the right.}
\end{figure}

Development of SW2GDML is ongoing. It can convert simple SOLIDWORKS designs and supports the following shapes and features:
board, cone, cylinder (full and partial), disk (full and partial), half-ellipsoid with a circular face, 
torus, cylindrical holes in parts, multiple coordinate systems in simple configurations, and repeated parts in linear patterns.
Figures~\ref{swbigtank}, \ref{swinnertank},  and \ref{swfulltank} show examples of simple models automatically converted from SOLIDWORKS to Geant4 by SW2GDML.
An example of GDML generated by SW2GDML can be found in the Appendix.

SW2GDML consists of about 3000 lines of C++ code developed with Microsoft Visual Studio and employs the SOLIDWORKS Application Program
Interface (API).  This API gives the converter direct access to the details of the model stored in SOLIDWORKS.
Conceptually, the code is divided into three modules: one to read out of SOLIDWORKS the information
about parts and their surfaces, one to associate surfaces together to form solids and calculate necessary coordinate transformations for them,
and one to write out the solids in GDML format. The reading module is the most complicated one, because the SOLIDWORKS API requires many complex
operations to extract and calculate the features of the parts.

SW2GDML is tied directly to SOLIDWORKS, and this tie is both an advantage and a disadvantage. It is able to very efficiently access all information
in the SOLIDWORKS model, but it cannot work at all with any other CAD program. To adapt the SW2GDML code to another CAD program would
be a very large development effort. Though the non-SOLIDWORKS-specific parts of the code could be probably be used with few changes,
a new CAD API would probably be
entirely different from the SOLIDWORKS API and thus require a complete re-development of the most complex part of SW2GDML.

\section{Future Development of SW2GDML}
SW2GDML cannot yet handle complex SOLIDWORKS models. Features not supported by SW2GDML may be omitted or may cause misplacement of
converted parts. SOLIDWORKS has a long history of continuous development
and enhancement, and it has become quite complex, with several different, parallel methods to perform each design task.
Implementing in SW2GDML support for a feature in SOLIDWORKS has to be done not only once but
several times because of the multiple approaches SOLIDWORKS uses to handle the same feature or capability.
This complexity presents a large challenge to the goal of fully supporting conversion of all possible 
SOLIDWORKS models. At present, the SW2GDML development plan is incremental, with the aim of adding
support for the most important or commonly used SOLIDWORKS features in a step-by-step fashion.

\begin{figure}[htpb]
\begin{minipage}{16pc}
\includegraphics[width=16pc,trim={3.0cm 0 3.0cm 0},clip,scale=0.08]{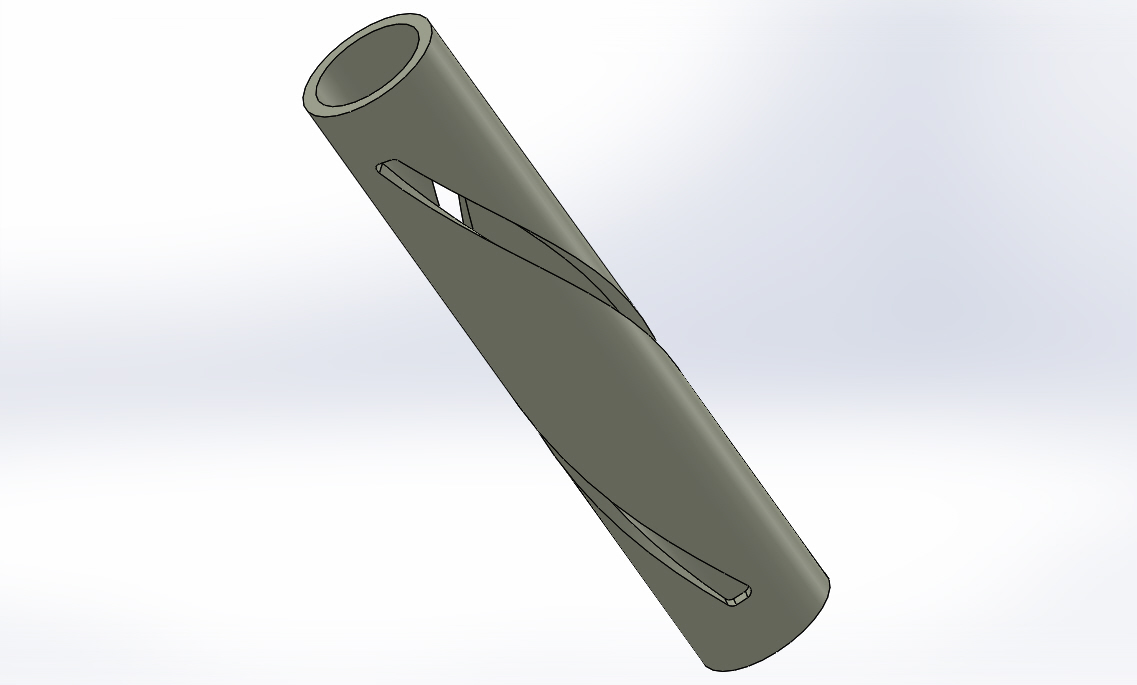}
\end{minipage}\hspace{2pc}%
\begin{minipage}{16pc}
\includegraphics[width=16pc,scale=0.1]{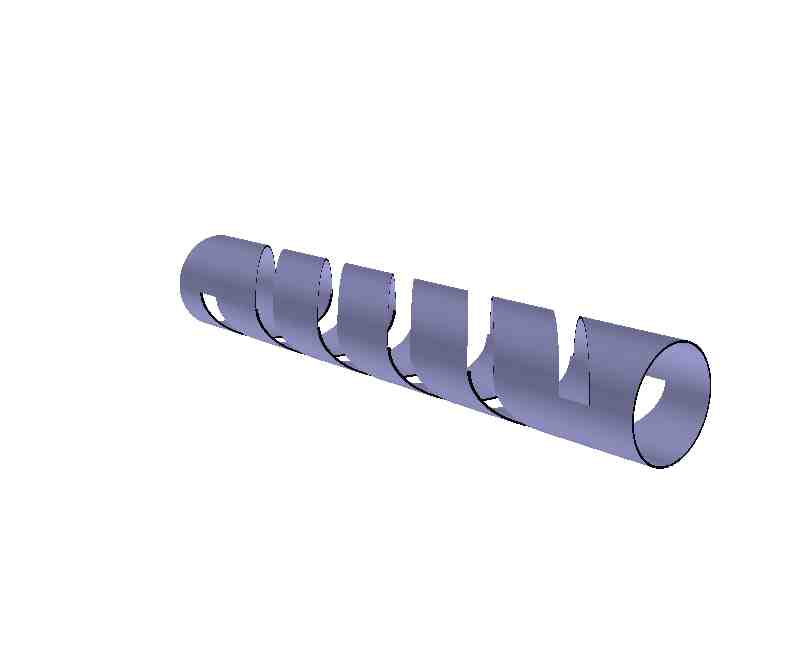}
\end{minipage} 
\caption{\label{swspiralcut} X-ray collimator model~\cite{dinca},
shown in SOLIDWORKS on the left. On the right is a manually created, proof-of-concept Geant4 model.}
\end{figure}

The current priorities for development are support for additional shapes, more complex combinations of coordinate systems, and special
patterns where a single part is specified to repeat around a circle. Figure~\ref{swspiralcut} shows 
an x-ray collimator design~\cite{dinca} with a spiral-cut cylinder shape whose conversion is currently under development.

\section{Conclusion}
SW2GDML is a tool to automatically convert designs in the SOLIDWORKS CAD program into GDML for importing into Geant4 for simulation. It aims
to facilitate rapid, iterative cycles of design, simulation, and validation. It currently
supports conversion of simple SOLIDWORKS models and is on a path of further development to support more SOLIDWORKS features.

\ack
We gratefully acknowledge the support for this work provided by grant \mbox{PHY-1104447} from the National Science Foundation.

\appendix
\section*{Appendix}

An abbreviated example of GDML generated by SW2GDML follows:

{\fontfamily{pcr}\selectfont
\begin{verbatim}

<materials>
  <material name="AISI_316_Stainless_Steel_Sheet_.SS.">
   <D value="8.0" unit="g/cm3"/>
   <fraction n="0.685" ref="Iron"/>
   <fraction n="0.17" ref="Chromium"/>
   <fraction n="0.12" ref="Nickel"/>
   <fraction n="0.025" ref="Molybdenum"/>
  </material>
 </materials>

<solids>
 <box name="WorldBox" x="10000.0" y="10000.0" z="10000.0"/>
 <cone name="cone1" z="0.687022" rmin1="0" rmin2="0" rmax1="0.126993"
  rmax2="0.0746125" deltaphi="TWOPI"/>
 <torus name="torus1" rtor="0.119" rmin="0" rmax="0.054" deltaphi="1.5708"/>
 <ellipsoid name="s-revolve1" ax="0.91395" by="0.91395" cz="0.460433"
  zcut1="0"/>
 <tube name="disk1" rmin="0" rmax="3.81476" deltaphi="6.28319" z="0.01"/>
 <tube name="cylinder1" z="5.96274" rmin="3.81" rmax="3.81476"
  deltaphi="6.28319"/>
 <tube name="cylinder12" z="0.342" rmin="0" rmax="0.054" deltaphi="6.28319"/>
 <subtraction name="subt1">
  <first ref="torus1"/>
  <second ref="cylinder12"/>
  <position name="pos1" x="-0.171" y="0.119" z="0"/>
  <rotation name="rot1" x="0" y="1.5708" z="0"/>
 </subtraction>
</solids>

<structure>
 <volume name="vol8">
  <materialref ref="AISI_316_Stainless_Steel_Sheet_.SS."/>
  <solidref ref="disk1"/>
 </volume>
 <volume name="World">
  <materialref ref="Air"/>
  <solidref ref="WorldBox"/>
  <physvol>
   <volumeref ref="vol8"/>
   <position name="pos11" x="0" y="0" z="5.93734"/>
   <rotation name="rot11" z="0" y="1.5708" x="0"/>
  </physvol>
 </volume>
</structure>

\end{verbatim}
}

\bibliography{sw2gdml.bib}

\end{document}